\documentclass[prl,twocolumn,showpacs,floatfix,amsmath,footinbib]{revtex4}

\usepackage{amssymb}
\usepackage{graphicx}
\usepackage{dsfont}
\usepackage{overpic}
\usepackage{times}
\usepackage{color}

\newcommand{\vect}[1]{\ensuremath{\mathbf{#1}}}

\newcommand{\dd}{\ensuremath{\mathrm{d}}}

\newcommand{\uvect}[1]{\ensuremath{\hat{\mathbf{#1}}}}
\newcommand{\expo}[1]{\ensuremath{\mathrm{e}^{#1}}}

\newcommand{\bsqks}{\ensuremath{b^2_{\vect{k} \uvect{s}}}}
\newcommand{\bsq}{\ensuremath{b^2_{\uvect{s}}}}

\makeatother

\begin{document}

\title{Anisotropy of spin relaxation in metals}
\author{Bernd Zimmermann}
\author{Phivos Mavropoulos}\email{Ph.Mavropoulos@fz-juelich.de}
\author{Swantje Heers}
\author{Nguyen H. Long}
\author{Stefan Bl\"{u}gel}
\author{Yuriy Mokrousov}
\affiliation{Peter Gr\"{u}nberg Institut and Institute for Advanced Simulation, Forschungszentrum J\"{u}lich and JARA, 52425 J\"{u}lich, Germany}
\date{\today}

\begin{abstract}

  The concept of anisotropy of spin relaxation in non-magnetic metals
  with respect to the spin direction of the injected electrons
  relative to the crystal orientation is introduced. The effect is
  related to an anisotropy of the Elliott-Yafet parameter, arising
  from a modulation of the decomposition of the spin-orbit Hamiltonian
  into spin-conserving and spin-flip terms as the spin quantization
  axis is varied. This anisotropy, reaching gigantic values for
  uniaxial transition-metals (e.g.~830\% for hcp Hf) as
  density-functional calculations show, is related to extended
  ``spin-flip hot areas'' on the Fermi surface created by the
  proximity of extended sheets of the surface, or ``spin-flip hot
  loops'' at the Brillouin zone boundary, and has no theoretical upper
  limit. Possible ways of measuring the effect as well as consequences
  in application are briefly outlined.

\end{abstract}



\maketitle

Spin-relaxation processes are of fundamental importance for the
realization of spintronic devices, which aim at utilizing the electron
spin degree of freedom for processing and transfer of information
\cite{Zutic.2004.rmp}. A non-equilibrium spin-distribution will
generally equilibrate and the corresponding information will be lost
on a timescale of the spin-relaxation time $T_1$.  The dominant
spin-relaxation mechanism in structure- and bulk-inversion symmetric,
non-magnetic metals is the Elliott-Yafet
mechanism~\cite{Elliott.1954,Yafet.1963}, which is due to scattering
of electrons at phonons or impurities. Owing to the presence of
spin-orbit coupling (SOC) in a solid, such scattering events will flip
the spin of an electron with a certain probability, which depends on
both the wavefunctions of the ideal crystal and the scattering
potential. An estimate of the ratio between momentum- and
spin-relaxation time, $T_\mathrm{p}$ and $T_1$, can be given 
in the diffusive regime in a
first approximation by neglecting the form of the scattering
potential. Within this \emph{Elliott approximation} one obtains that $T_\mathrm{p} / T_1 \approx 4 b^2$,
where $b^2$ is the \emph{Elliott-Yafet (or spin-mixing) parameter}, which we
define below  and which, being a property of the ideal crystal
only~\cite{Elliott.1954,Yafet.1963}, is related to the spin-orbit
coupling strength in a material. Thus, $b^2$ has since long been accepted as a
measure of spin relaxation, provided that some mechanism of momentum
relaxation is present. However, what has not been analyzed so far on a
theoretical basis is the anisotropy of $b^2$, leading to a certain
anisotropy of spin relaxation \cite{foot1}.

We put our notion of anisotropy in more precise terms by the following
considerations. In an experiment measuring spin relaxation the
injected electrons are always characterized by the axis of spin
polarization which is determined e.g. by the external magnetic field
in conduction-electron spin resonance or by the direction of
magnetization of ferromagnetic leads in spin-injection or
giant-magnetoresistance experiments. For our purposes we call this
direction the \emph{spin quantization axis} (SQA). As it turns out,
the SQA relative to the crystal lattice matters for the value of
$b^2$, and not just by a little. This gives us the notion of
anisotropy of the Elliott-Yafet parameter and consequently of spin
relaxation: It will thus make a difference if spins are injected into
a metal from a ferromagnet whose magnetization is normal to the
interface, compared to being parallel.
Since the spin population decays exponentially with
respect to the distance from the injection point, with $T_1$ entering
in the exponent, the anisotropy can make a difference between e.g. a
well-operating and a defective giant-magnetoresistance junction. One
realizes also that we are faced with an anisotropy of an irreversible
process, i.e., of the entropy production during relaxation. This can
have far-reaching consequences, for example in the spin-entropy
induced Peltier cooling in nanojunctions \cite{Yoshida07}. In this
respect this anisotropy is fundamentally different from other, known
SOC effects, such as the magnetocrystalline anisotropy energy of
ferromagnets or the anisotropic magnetoresistance which arise from
small changes of the band energies depending on the magnetization
direction. 
From a band-structure point of view, in a nonmagnetic metal the choice
of the SQA does not influence the band energies, but it manifests
itself through matrix elements determining the orbital and spin
character of the Bloch states. This, for example, leads 
to large changes of the spin Hall conductivity in non-cubic transition 
metals when the direction of the spin current's polarization  
is varied~\cite{Freimuth.2010.prl}. One also expects an
anisotropy in the spin susceptibility, since the spin-mixing parameter
of a state is directly related to its response to a Zeemann field.

In this Letter, we investigate the anisotropy of the Elliott-Yafet
parameter in metals. For this purpose we employ density functional
theory, which has been successfully applied in the past to calculate
the spin-mixing parameter in various metals~\cite{Gradhand.2009.prb,
  Steiauf.2009.prb}. Our main finding is that in non-cubic transition
metals, or generally metallic systems of lowered symmetry, e.g.~in the
hcp structure, the anisotropy of the Elliott-Yafet parameter can
be gigantic. We also demonstrate that in metals with cubic symmetry,
such as~e.g.~bcc tungsten or fcc gold, this anisotropy is much
smaller, although it can still reach large values. Moreover, we
analyze the Fermi surface properties of the spin-mixing anisotropy and
provide simple arguments for a microscopic understanding of our
results. Our choice of materials ($5d$ metals) is based on their similar spin-orbit
strength, but different crystal structures, which at the end brings
about anisotropy values differing by orders of magnitude.

\begin{figure*}[htb]
   \includegraphics[width=170mm]{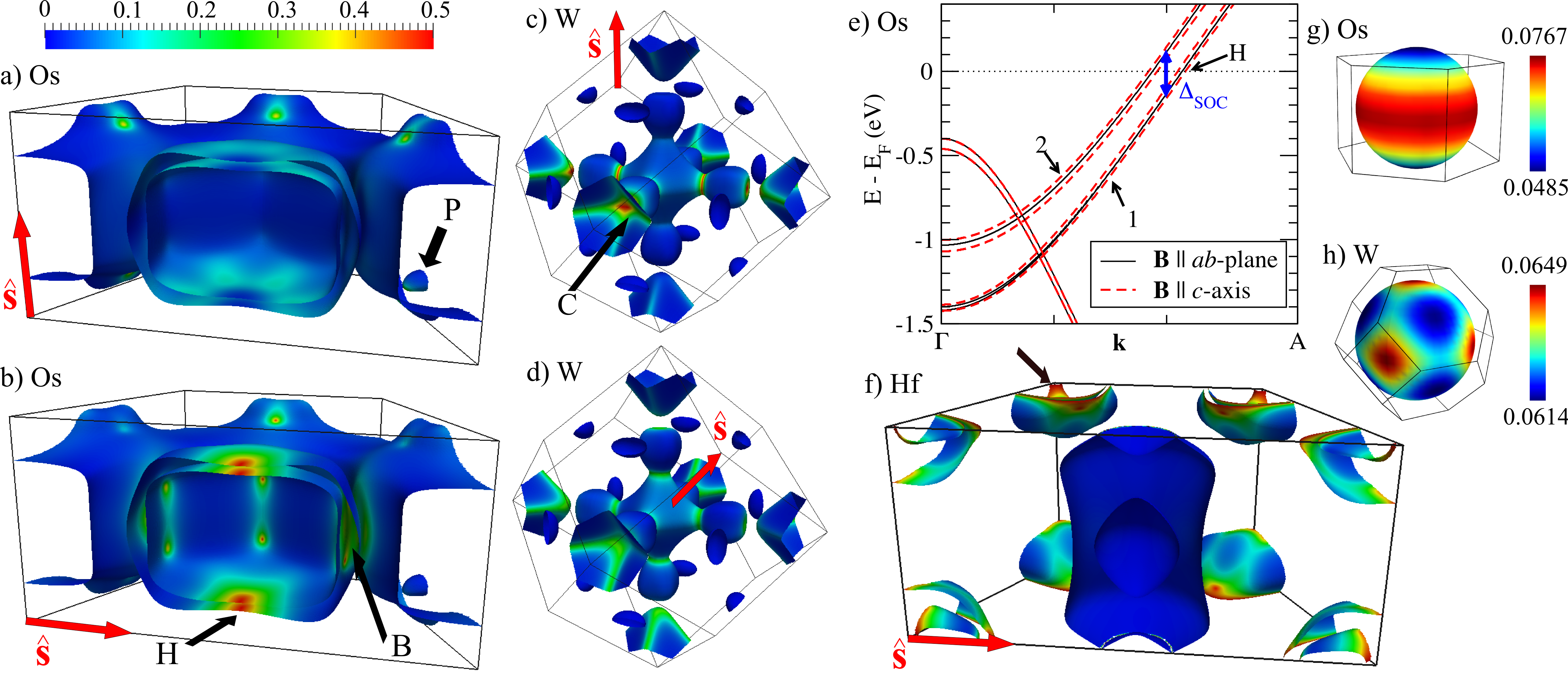}
   \caption{(color online) Fermi surfaces of Os (a-b), W (c-d) and Hf
     (f). For an illustration of the nested sheets of Os and Hf, only
     half of the Fermi surface is shown. The Elliott-Yafet parameter
     $\bsqks$ of Os is shown in terms of a color code on the Fermi
     surface with the SQA $\uvect{s}$ [red arrows at the left-lower
     corners] along the $c$-axis (a) and in the $ab$-plane (b). The
     splitting introduced by spin-orbit coupling and by a Zeemann-like
     field is shown in the band structure of Os (e) along the
     $\Gamma$-$A$ direction (BZ center to hexagonal-face
     center). Analogously, $\bsqks$ for $\uvect{s}$ along $[001]$ and
     $[111]$ in W is shown in (c) and (d), respectively. In (g) and
     (h), the integrated Elliott-Yafet parameter $\bsq$ is shown as
     function of the SQA direction for Os and W, respectively (notice
     that in (g,h) the color-scale is different than in the
     Fermi-surface plots). The averaged values of $b^2_{\uvect{s}}$
     over all directions of $\uvect{s}$, corresponding to
     polycrystalline samples, are 0.0666 for Os and 0.0627 for W. (f)
     Fermi surface of Hf with the value of $\bsqks$ shown as
     colorcode, where $\uvect{s}$ is parallel to the $ab$-plane. An
     arrow indicates one of the spin-flip hot loops, clearly visible
     in red on the hexagonal face of the BZ. The hot loops vanish when
     $\uvect{s}$ is rotated to the $c$-axis, resulting in an
     anisotropy of 830\% of the Elliott-Yafet parameter.}
  \label{fig:fermisurfaces}
\end{figure*}

The Elliott-Yafet theory is based on the observation that the
spin-orbit coupling of the lattice ions causes the Bloch eigenstates
to be a superposition of spin-up and spin-down character. This
superposition is often called \emph{spin mixing}. The
spin-orbit Hamiltonian can be divided into a spin-conserving
$\xi(LS)_\parallel$ and a spin-flip part $\xi(LS)^{\uparrow
  \downarrow}$, given respectively by the first and second parts of
the r.h.s.\ of the following expression:
\begin{equation}
  \xi \vect{L} \cdot \vect{S} = \xi L_{\uvect{s}} S_{\uvect{s}} + \frac{1}{2}\xi \left( L_{\uvect{s}}^{+} S_{\uvect{s}}^{-} + L_{\uvect{s}}^{-} S_{\uvect{s}}^{+} \right). \label{Eq:spinorbit}
\end{equation}
Here, $\xi$ is the spin-orbit coupling strength, $\uvect{s}$ is a unit
vector in the direction of the spin- (and generally angular momentum-)
quantization axis, $\vect{L}$ and
$\vect{S}=\frac{\hbar}{2}\boldsymbol{\sigma}$ are the orbital and spin
angular momentum operators respectively, $L_{\uvect{s}} = \vect{L}
\cdot \uvect{s}$, $S_{\uvect{s}} = \vect{S} \cdot \uvect{s}$, and
$L_{\uvect{s}}^{\pm}$ and $S_{\uvect{s}}^{\pm}$ are the corresponding
raising and lowering operators for angular momentum and spin in the
reference frame specified by the direction vector $\uvect{s}$
(defining the SQA). It is clear that the dot product $\vect{L} \cdot
\vect{S}$ is independent of $\uvect{s}$, leaving the eigenenergies of
the Hamiltonian invariant. However, the spin-conserving and spin-flip
parts separately depend on the choice of the SQA.

The time-reversal and space-inversion symmetries imply that the
eigenenergies of the system at any Bloch momentum $\vect{k}$ are at
least two-fold degenerate, with the corresponding states taking the
form \cite{Elliott.1954}
\begin{eqnarray}
{\Psi}_{\vect{k} \uvect{s}}^{+} (\vect{r}) &=& \left[ a_{\vect{k}  \uvect{s}} (\vect{r}) ~ \lvert \uparrow \rangle_{\uvect{s}} + b_{\vect{k}  \uvect{s}} (\vect{r}) ~ \lvert \downarrow \rangle_{\uvect{s}} \right] ~ \expo{i \vect{k} \cdot \vect{r}} ~, \\
{\Psi}_{\vect{k} \uvect{s}}^{-} (\vect{r}) &=& \left[ a_{-\vect{k} \uvect{s}}^{*} (\vect{r}) ~ \lvert \downarrow \rangle_{\uvect{s}} - b_{-\vect{k} \uvect{s}}^{*} (\vect{r}) ~ \lvert \uparrow \rangle_{\uvect{s}} \right] ~ \expo{i \vect{k} \cdot \vect{r}}~.
\end{eqnarray}
The two spin states $\lvert \uparrow \rangle_{\uvect{s}}$ and $\lvert
\downarrow \rangle_{\uvect{s}}$ are eigenstates of $\vect{S} \cdot
\uvect{s}$, e.g.~ if $\uvect{s}\parallel z$, $\lvert \uparrow
\rangle_{z}$ and $\lvert \downarrow \rangle_{z}$ are the eigenstates
of the $S_z$ operator. The functions $a_{\vect{k}
  \uvect{s}}(\vect{r})$ and $b_{\vect{k} \uvect{s}}(\vect{r})$ exhibit
the periodicity of the crystal lattice. We define $\bsqks$ as the unit
cell (u.c.) integral $\int_{\mathrm{u.c.}}{ \dd^3 r \, \lvert
  b_{\vect{k} \uvect{s}} (\vect{r}) \rvert^2 }$.

For fixed direction $\uvect{s}$, the degenerate ${\Psi}_{\vect{k}
  \uvect{s}}^{+}$ and ${\Psi}_{\vect{k} \uvect{s}}^{-}$ states [and
the corresponding $ a_{\vect{k} \uvect{s}} (\vect{r})$ and
$b_{\vect{k} \uvect{s}} (\vect{r})$] can be chosen, by linear
combination, such that the spin-expectation value $\langle
S_{\uvect{s}}\rangle_{\vect{k}} = \langle {\Psi}_{\vect{k}
  \uvect{s}}^{+} \rvert S_{\uvect{s}} \lvert {\Psi}_{\vect{k}
  \uvect{s}}^{+} \rangle$ is maximal. The spin mixing parameter is
then given by $\bsqks = 1/2 - \langle
S_{\uvect{s}}\rangle_{\vect{k}}/\hbar$, and is usually small, due to
the weakness of the SOC. In this case the Bloch states are of nearly
pure spin character. However, at special ``spin-flip hot spot'' points in
the Brillouin zone (BZ),~e.g.~accidental degeneracies, BZ boundaries
or other high symmetry points~\cite{Fabian.1998.prl,Fabian.1999.prl},
$\bsqks$ may increase significantly up to $\frac{1}{2}$, which
corresponds to the case of fully spin-mixed states. The Fermi-surface
(FS) averaged spin-mixing, or Elliott-Yafet, parameter is given by
\begin{equation}
  \bsq = \frac{1}{n(E_F)} ~ \frac{1}{\hbar} \, \int_\mathrm{FS}{  \frac{  \bsqks   }{ \lvert\vect{v}(\vect{k})\rvert } ~ d^2 k }~, \label{integral}
\end{equation}
where $\vect{v}(\vect{k})$ is the Fermi velocity. The normalization by
the density of states at the Fermi level, $n(E_F) = 1/\hbar \,
\int_\mathrm{FS}{ \lvert \vect{v}(\vect{k}) \rvert^{-1} ~ \dd^2 k }$,
ensures that $ 0\leq \bsq \leq \frac{1}{2}$.

We use density functional theory in the local density approximation
\cite{Vosko.1980.CanJPhys} to calculate the electronic structure of
metals considered in the following. For the self-consistent
calculations we employ the Korringa-Kohn-Rostoker (KKR) Green-function
method \cite{JuelichMuenichCode} in the atomic sphere approximation
and solve the Dirac equation with an angular-momentum expansion up to
$\ell_\mathrm{max}=4$. 
We choose a grid of at least 200 $\vect{k}$-points along each direction in
the full Brillouin zone, resulting in about $10^7$ Fermi-surface
points. We follow the procedure described in
Ref.~[\onlinecite{Heers.PhD}] to maximize the spin component
$S_{\vect{k} \uvect{s}}$ at the Fermi-surface points. The details of determination 
of the Fermi surface and corresponding Fermi surface integration will be 
published elsewhere.

First, we turn to hcp osmium, which exhibits a uniaxial crystal
structure, and discuss the results in detail.  The Fermi surface of Os
presented in Fig.~\ref{fig:fermisurfaces}(a-b) consists of two nested
sheets, a surrounding surface crossing the BZ boundary and little hole
pockets ``P''. 
Analyzing the distribution of the spin mixing parameter $\bsqks$ on
the Fermi surface, we observe a strong dependence on the SQA, evident
from comparing Figs.~\ref{fig:fermisurfaces}(a) and (b). For $\uvect{s}$ along the $c$-axis
of the crystal [Fig.~\ref{fig:fermisurfaces}(a)], the spin mixing is
relatively uniform ($\bsqks \approx 0.05$) for large areas of the
Fermi surface, reaching larger values near the pockets. However, this
picture changes drastically when $\uvect{s}$ is parallel to the
$ab$-plane [Fig.~\ref{fig:fermisurfaces}(b)]. In this case, areas
with full spin mixing (red, $\bsqks \approx 0.5$) are prominent, most
clearly visible at the caps of the two nested Fermi-surface sheets
(indicated by ``H''). Additionally, large areas with smaller, but
still strong spin mixing ($\bsqks \approx 0.3$) are visible,~e.g.~in
the area denoted by ``B''. Overall, for the two considered cases
there is a strong qualitative difference in the $\vect{k}$-dependent
spin-mixing parameter $\bsqks$.

As for the Fermi-surface averaged $\bsq$, we find values of $4.85
\times 10^{-2}$ and $7.69 \times 10^{-2}$ for $\uvect{s}$ along the
$c$-axis and parallel to the $ab$-plane, respectively, yielding thus a
gigantic anisotropy of the Elliott-Yafet parameter, defined as
$\mathcal{A} = \left[ \max_{\uvect{s}}(\bsq) - \min_{\uvect{s}}(\bsq)
\right]/ \min_{\uvect{s}}(\bsq)$, of 59\%. The anisotropy
with respect to rotations of the SQA within the
$ab$-plane is, on the other hand, negligible. These two limiting cases
are contained in Fig.~\ref{fig:fermisurfaces}(g), in which the value
of $\bsq$ is shown as a function of all possible directions of
$\uvect{s}$ on the unit sphere. The absent (or very small) anisotropy
within the $ab$-plane is reflected in the rotationally invariant
color-scale around the $c$-axis, as opposed to the large difference
between the $ab$-plane and the $c$-axis.
A detailed analysis reveals that the largest contribution to $\mathcal{A}$
comes from the somewhat extended areas with high or intermediate values of  $\bsqks$
between 0.15 and 0.35, which we refer to as  ``spin-flip hot areas'' in the following.

We delve into the anisotropy of the spin-mixing at the area around H
by means of a numerical experiment. Analyzing the band structure of Os
along the high symmetry line from the center of the BZ
($\Gamma$-point) through the H-point to the center of the hexagonal
face ($A$-point), in Fig.~\ref{fig:fermisurfaces}(e), we see two bands
crossing the Fermi level at H (full black lines), representing the two
nested Fermi-surface sheets of Fig.~\ref{fig:fermisurfaces}(a-b). The
splitting $\Delta_{\rm SOC}$ between these two bands is due to SOC, as
we have verified by the fact that they fall on top of each other when
the spin-orbit coupling strength is scaled down (not shown). Each band
is twofold degenerate due to time-reversal and inversion
symmetry. This remanent degeneracy can be lifted by applying a small
Zeeman-like field $\vect{B}$ coupling to the Bloch states via a term
$\boldsymbol{\sigma} \cdot \vect{B}$, which breaks the time-reversal
symmetry and sets the SQA in the direction of $\vect{B}$. We choose a
weak $\vect{B}$-field with a magntiude of 40~meV and vary its
direction. In Fig.~\ref{fig:fermisurfaces}(e) we clearly observe a
splitting of bands ``1'' and ``2'' for $\vect{B}$ along the $c$-axis
(dashed red lines). However, for $\vect{B}$ in the $ab$-plane, the
degenerate pairs ``1'' and ``2'' do not split (solid black lines). We
can relate this result to our findings for $\bsqks$ by employing
perturbation theory arguments: in first order, the energy shift of a
state in presence of a small $\vect{B}$-field is proportional to the
state's spin polarization. This numerical experiment shows that in the
case that $\vect{B}\parallel ab$-plane the states are fully spin
mixed, just as found by the calculation of
$b^2_{\vect{k};\hat{\vect{s}}\parallel ab}$. It also reveals an
anisotropy of the susceptibility that follows from the anisotropy of
the spin-mixing parameter.

We are now in a position to give a simple line of arguments
demonstrating the microscopic mechanism that leads to a large
anisotropy in general. As a working example we use the calculated
giant anisotropy at and in the vicinity of the H-point in Os.  Important in
the setup is the presence of a degeneracy or near-degeneracy at $E_F$,
here at the point H, of Bloch wavefunctions ${\Psi}^{1,2}_{\vect{k}}$,
which in a tight-binding picture we represent as
${\Psi}^{1,2}_{\vect{k}}=\sum_{lm} c^{1,2}_{\vect{k}lm}
|l,m\rangle$. Here, $|l,m\rangle$ are eigenstates of
$\vect{L}_{\uvect{s}}$ for a given SQA $\uvect{s}$ in the crystal,
e.g. the $z$ axis.  The requirement for a large anisotropy is met if
the matrix elements of the spin-flip SOC operator $\xi(LS)^{\uparrow
  \downarrow}$ between ${\Psi}^{1}_{\vect{k}}$ and
${\Psi}^{2}_{\vect{k}}$ vanish. This occurs if the expansions above
\emph{exclude} terms with $l_1=l_2$ and $|m_1-m_2| = 1$, where
$|l_1,m_1\rangle$ contributes to ${\Psi}^{1}_{\vect{k}}$ and
$|l_2,m_2\rangle$ to ${\Psi}^{2}_{\vect{k}}$. Here, concretely, we
have found that only the orbitals $|d,-1\rangle$ and $|d,+1\rangle$
contribute to the bands 1 and 2 at H. These are superimposed to form
the oriented orbitals $d_{xz}$ and $d_{yz}$ which are the actual
eigenstates of the crystal field. Then the lifting of degeneracy is
only due to the spin-conserving SOC, $\xi(LS)_{\parallel}$ which
causes no spin-mixing.

Suppose now that the SQA (and together with it the axis for
quantization of orbital angular momentum) is rotated around $y$, from
$z$ to $\bar{z}=x$. In the new frame ($\bar{x}\bar{y}\bar{z}$) we
denote the orbital functions and angular momentum indices with an
overline. The oriented orbitals have a new resolution with respect to
the new axes. For example, $d_{yz}$ becomes $d_{\bar{x}\bar{y}}$
having a projection on $|\overline{d,\pm 2}\rangle$, while $d_{xz}$
becomes $d_{\bar{x}\bar{z}}$ having a projection on $|\overline{d,\pm
  1}\rangle$. As a result, the expansions of ${\Psi}^{1,2}_{\vect{k}}$
with respect to the new frame \emph{include} orbitals with $l_1=l_2$
and $|\overline{m}_1-\overline{m}_2|= 1$ at the same energy, allowing
for non-zero matrix elements of the spin-flip part of the SOC.  For
the SQA along $z$, the system is ``protected'' against large-amplitude
spin-flip transitions, while for the SQA along $x$ or $y$ spin-flip
transitions are favored. What we have demonstrated here is that the
matrix elements of the spin conserving and spin-flip part of the SOC
can depend so strongly on the SQA that the spin-flip vanishes in one
direction while it is maximal in another. For example, while in the
original frame the SOC-induced band splitting at H arises from
$\xi(LS)_\parallel$, in the new frame the exact same splitting arises
from $\xi(LS)^{\uparrow \downarrow}$, leading to a spin mixing of
$\frac{1}{2}$. It is, therefore, the direction of the SQA that
dictates which part of the Hamiltonian causes most of the splitting,
even though the sum of the two contributions is independent of the
direction of the SQA. Other states, being higher or lower in energy
due to the crystal field splitting, play only a small role in the
final result.

The mechanism for large anisotropy of the spin-mixing
parameter described above is of course not only specific to the $d$
states of Os, but it is also responsible for large values of
$\mathcal{A}$ that we find for hcp Lu (200\%), hcp Re (88\%) and hcp
Hf (830\%). Particularly in hcp metals there is a special symmetry at
the hexagonal face ot the Brillouin zone that is lifted only by the
SOC \cite{Ashcroft}. Thus, whenever the Fermi surface of an hcp metal
happens to cut through the hexagonal face, the resulting contour can
obtain full spin-mixing depending on the SQA, as shown in
Fig.~\ref{fig:fermisurfaces}(f) for Hf. These looplike contours, or
``spin-flip hot loops,'' are a source of extremely high
anisotropy. The Fermi surfaces of Lu, Re and Hf for example contain
such loops, but the one of Os does not, since it does not cut through
the hexagonal face.  

We further observe that the magnitude of the effect can be strongly
enhanced by the large extension of the two near-degenerate, parallel
sheets of the Fermi surface, resulting in a spin-flip hot area around
the point of near degeneracy. E.g.~in Os we obtain a hot area around
point H instead of a single hot spot at H, while the position and
topology of such area generally depends on the Fermi surface and
electronic structure of the transition metal. In addition, the reduced
symmetry helps: if the crystal had cubic symmetry, then upon change of
the SQA from $z$ to $x$ the effects at rotationally equivalent parts
of the Fermi surface would mutually cancel.

We now turn to tungsten, which has a bcc lattice structure. When
$\uvect{s} \parallel [001]$, $\bsqks$ exhibits hot spots in directions
perpendicular to $\uvect{s}$ (denoted by ``C'')
[Fig.~\ref{fig:fermisurfaces}(c)], but not at the rotationally
equivalent points along the $z$-axis, following the formation scenario
similar to that at the H-point in Os.  Additionally, many states with
smaller spin mixing ($0.2 < \bsqks < 0.3$) are present at the Fermi
surface, leading to $\bsq = 6.49 \times 10^{-2}$. For SQA along
another high symmetry direction of the lattice, $\uvect{s} \parallel
[111]$ in Fig.~\ref{fig:fermisurfaces}(d), the intensity at the point
C is reduced, but a large area with smaller spin mixing is clearly
present, resulting in $\bsq = 6.14 \times 10^{-2}$. For SQA along
$[110]$, we find $\bsq = 6.26 \times 10^{-2}$. This leads to an
anisotropy $\mathcal{A}=6\%$, which is still large but one order of
magnitude smaller than in hcp Os. In fact we also predict rather
small anisotropy in other cubic transition metals, such as Ir (1\%) and 
Pt (0.4\%). This observation is similar to the dependence of the magnetocrystalline 
anisotropy energy and anisotropy of the intrinsic anomalous Hall conductivity
\cite{Zener.1954,Roman.2011} on the symmetry of the lattice in
ferromagnetic crystals: the cubic crystal exhibits a fourfold
rotational axis, causing SOC to contribute to $\mathcal{A}$ in fourth
order. In the uniaxial hcp structure, an axis perpendicular to the
$c$-axis is only twofold, leading SOC to enter $\mathcal{A}$ in second
order. 
Generally we expect the integrated value $\bsq$ to
exhibit the full point-group symmetry of the lattice (evident in Fig.~\ref{fig:fermisurfaces}(g) for Os and (h) for W), even if the map
of $\bsqks$ has a lower symmetry. 
The comparatively large anisotropy value
in W is partly a consequence of the $d$-states, which yield
a strong directional anisotropy of the Fermi surface. In contrast to
this, the Fermi surface of fcc gold consists of $s$-like states and
can be regarded as almost spherical. For the Elliott-Yafet parameter
in Au, we find $\bsq \approx 3.25 \times 10^{-2}$,
i.e.~the same order of magnitude as in W and Os, but the anisotropy is 
only 0.1\%.

A measurement of the
anisotropy requires samples with preferential crystalline orientation
and a rotation of the magnetic field direction in an electron spin
resonance experiment or a rotation of the magnetization direction of
the ferromagnetic injector in a spin-injection experiment.
In reality, the precise scattering mechanism will
naturally affect the result. Nevertheless our results should be
plainly measurable even if the exact value can deviate somewhat from
the one of the Elliott approximation. Perhaps the anisotropy, stemming
from the band structure, can be most conveniently measured as a
function of the temperature invoking phonon scattering, in order to
circumvent any possible extrinsic anisotropy arising from the electronic
structure of defects in the system. 

In conclusion, spin relaxation in metals can strongly depend on the
orientation of the injected-electron spin axis due to a corresponding
anisotropy of the Elliott-Yafet parameter. The anisotropy is expected
to be largest in non-cubic crystals and in the presence of extended
nested Fermi-surface sheets that are almost degenerate, resulting in
extended spin hot areas or hot loops instead of singular spin hot
spots; especially critical are cases where the splitting between 
nested sheets is caused primarily by the spin-orbit coupling. Since
there is no theoretical limit on the area of the nested sheets in this
scenario, the anisotropy can in principle exceed the large values
calculated here and is an effect worthwhile investigating on a number
of metals and ordered alloys. Possible symmetry reduction by the
Fermi-surface nesting, e.g. charge-density waves, would be of long
wavelength because of the smallness of the $\vect{k}$-vector
connecting the split Fermi-surface sheets, therefore we expect a small
reduction but not a fundamental change by such effects. Temperature or
moderate alloying can blur out the sharpness of the bands, but if the
band splitting is on the order of 0.2$-$0.3~eV (approx. 2000$-$3000
Kelvin), and the bandwidth of the critical bands on the order of 1$-$2
eV, as we found in the case of Os, then the blurring of the bands will
not be sufficient to mask the anisotropy. 
We furthermore
expect that anisotropy effects should also be present in metallic
alloys, heterostructures or ultrathin films, which we leave for future
work.

We are indebted to J. Fabian for an introduction to the field and for
discussions. We are also indebted to R.~Zeller and P.~H.~Dederichs for
their invaluable help in the KKR formalism and to G.~Bihlmayer,
D.V.~Fedorov and P.~Zahn for discussions. We acknowledge funding under
project MO 1731/3-1 and the Priority Programme SPP-1538 ``Spin Caloric
Transport'' of the Deutsche Forschungsgemeinschaft, the
Helmholtz-Gemeinschaft Young Inverstigators programme VH-NG-513, as
well as computing time at the J\"{u}lich Supercomputing Centre.


\end{document}